\documentclass[12pt]{article}
\usepackage{epsfig}
\usepackage{amsfonts}
\usepackage{amsopn}
\usepackage{amsmath}
\usepackage{verbatim,amsthm}

\title
{
\vskip-50 pt
\begin{flushright}
\normalsize\rm NORDITA-2012-64
\end{flushright}
\vskip 20 pt
Laplace-Beltrami operator and exact solutions for branes 
}
\author{
 A. A. Zheltukhin $^{a,b}$\thanks{e-mail: aaz@physto.se}  \\ \\
$^a$ Kharkov Institute of Physics and Technology, \\
1, Akademicheskaya St., Kharkov, 61108, Ukraine \\  
$^b$ Nordita\\
Royal Institute of Technology and Stockholm University\\
Roslagstullsbacken 23, SE-106 91 Stockholm, Sweden
}

\date{}

\begin{document}

\maketitle

\begin{abstract}
Proposed is a new approach to finding exact solutions of 
nonlinear $p$-brane equations in $D$-dimensional Minkowski space 
based on the use of various initial value constraints.
It is shown that the constraints $\Delta^{(p)}\vec{x}=0$ and 
$\Delta^{(p)}\vec{x}=-\Lambda(t,\sigma^r)\vec{x}$ give two sets of 
 exact solutions.

\end{abstract}

\section{Introduction}

Branes are fundamental constituents of string theory \cite{M1},
 but not so much is known about their internal structure 
 encoded in the nonlinear PDEs [2-14]. 
\nocite{tucker, hoppe1, BST, DHIS, FI, WHN, 
Z_0, WLN, BZ_0, hoppe2, hoppe3, Pol, UZ} 
As a result, the emphasis in the investigations is shifted 
to exploring various particular solutions and the physics 
based on them.  There is a progress 
in search for spinning membranes ($p=2$) with spherical/toroidal topology 
embedded in flat and  curved $AdS_p\times S^q$ backgrounds 
(see e.g.  [15-20] \nocite{CT, KY, HN, AFP, hoppe6}).
Extension of these results to the case $p=3$ and complexified backgrounds 
with symmetry groups such as $SU(n)\times SU(m)\times SU(k)$ 
was done in \cite{AF}, where radial stability of three-branes was 
established. Analysis of spinning branes with higher $p$, as well as 
 finding other particular solutions of brane equations is an open problem.
On this way an important observation was done by Hoppe in \cite{JU1}, 
where the $U(1)$-invariant anzats reducing the membrane equations in $D=5$ 
Minkowski space to the system of two-dimensional nonlinear equations 
was proposed. The particular solutions of the Hoppe equations which 
describe collapsing or spinning flat tori in $D=5$ were 
found in \cite{TZ} and their connection with the geometric 
approach, Abel and pendulum differential equations was established in \cite{ZT}. 
The extension of the membrane anzats describing the Abelian $U(1)^p$ 
invariant $p$-branes revealed exact hyperelliptic solutions for 
flat $p$-tori embedded into $D=(2p+1)$-dimensional Minkowski 
space \cite{Znpb}. Exact solutions corresponding to spinning $p$-branes 
in $D=(2p+1)$-dimensional Minkowski space were found in \cite{IJGM}.  

Here we make an attempt to understand the above-mentioned exact $p$-brane
 solutions on the base of a general approach which allows to find new 
exact solutions. The approach uses the wave 
representation of $p$-brane equations on the $(p+1)$-dimensional worldvolume 
$\Sigma_{p+1}$. In the orthogonal gauge these wave equations are reduced 
to the ones including Laplace-Beltrami operator $\Delta^{(p)}$ on the 
  hypersurface $\Sigma_{p}$. We propose to classify the brane solutions 
exploring various initial value constraints imposed on $\Delta^{(p)}\vec{x}$.
 We show that the harmonicity constraints $\Delta^{(p)}\vec{x}=0$ pick up 
the solutions describing spinning $p$-branes which include the spinning 
anzats \cite{IJGM} in the case  $D=2p+1$. These solutions include the
 infinite $p$-branes with the shape of hyperplanes which are reduced to 
$p$-dimensional domain walls with the constant brane energy density in the static 
limit. Found also are periodic solutions describing closed spinning 
folded $p$-branes with a singular metric, which generalize the folded 
string solutions \cite{ZYaF}, \cite{NNK} to the case of $p$-branes. 
The effect of the  formation of singularities for closed strings and 
membranes was also discussed in \cite{Hops}. 

Further, the present paper reveals that the harmonicity constraints 
$\Delta^{(p)}\vec{x}=-\Lambda(t,\sigma^r)\vec{x}$ select the exact solutions 
with $\Lambda=\frac{p}{\vec{R}^2(t)}$ describing closed 
$p$-branes with their hypersurface $\Sigma_{p}$ lying on the collapsing 
sphere $S^{D-2}$ with the time-dependent radius equal to $\sqrt{\vec{R}^2}$. 
The nonlinear equation for $R(t)$ turns out to be exactly solvable 
for any dimension $D$ of the  Minkowski space and results in hyperelliptic 
functions. In the case  $D=2p+1$ these solutions 
are reduced to the degenerate anzats \cite{Znpb} with all equal radii of 
the corresponding $p$-tori. The presence of such collapsing solutions generated 
by the deformed harmonicity constraint is a common property  of closed membranes 
and $p$-branes independent of the Minkowski space dimension $D\geq p+1$.

\section{Worldvolume wave equations for branes}

The Dirac action for a p-brane without boundaries
is defined  by the integral 
\begin{equation}\label{1}
S=T\int \sqrt{|G|}d^{p+1}\xi,
\end{equation}
in the dimensionless worldvolume parameters $\xi^{\alpha}$ ($\alpha=0,\ldots,p$). 
       The components $x^{m}=(t,\vec{x})$ of the brane world vector in
 the D-dimensional Minkowski space with the signature $\eta_{mn}=(+,-,\ldots,-)$
have the dimension of length, and the dimension of tension $T$ is $L^{-(p+1)}$.
The induced metric $G_{\alpha \beta}:=\partial_{\alpha} x_{m}\partial_{\beta} x^{m}$ 
is presented in $S$ by its determinant $G$.

 After splitting the parameters $\xi^{\alpha}:=(\tau,\sigma^r)$ 
the Euler-Lagrange equations and
$(p+1)$ primary constraints generated by $S$ take the form
\begin{equation}\label{5}
\partial_{\tau}{\mathcal{P}}^{m}=-T\partial_r(\sqrt{|G|}G^{r\alpha}\partial_{\alpha}x^{m}),
 \ \ \
\mathcal{P}^{m}=T\sqrt{|G|}G^{\tau \beta}\partial_{\beta} x^{m},
\end{equation}
\begin{equation}
\tilde{T}_{r}:=\mathcal{P}^{m} \partial_{r} x_{m} \approx 0, \ \ \ \
\tilde{U}:=\mathcal{P}^{m}\mathcal{P}_{m}-T^{2}|\det G_{rs}| \approx  0, \label{6}
\end{equation}
where $\mathcal{P}^{m}$ is the energy-momentum density of the brane.

It is convenient to use the orthogonal gauge simplifying the metric $G_{\alpha\beta}$  
\begin{eqnarray}\label{7}
L\tau=x^0\equiv t, \ \ \ \ G_{\tau r}= -L(\dot{\vec{x}} \cdot \partial_r \vec{x})=0, \\
 g_{rs}:=\partial_r \vec{x} \cdot \partial_s \vec{x}, \ \ \ \
G_{\alpha\beta}=\left( \begin{array}{cc}
                       L^{2}(1- {\dot{\vec{x}}}^2)& 0    \\
                         0 & -g_{rs}
                              \end{array} \right)
                               \nonumber
\end{eqnarray}
with $\dot{\vec{x}}:=\partial_{t}\vec{x}= L^{-1}\partial_{\tau}\vec{x}$.
As a result, the constraint $\tilde{U}$ (\ref{6}) represents $\mathcal{P}_0$ as
\begin{equation}\label{9}
\mathcal{P}_0=\sqrt{\vec{\mathcal{P}}^2+T^{2}|g|}, \ \ \ \
g=\det(g_{rs})
\end{equation}
and it becomes the Hamiltonian density $\mathcal{H}_0$ of the p-brane
since $\dot{\mathcal{P}}_0=0$ in view of Eq.(\ref{5}).
Using the definition of $\mathcal{P}_0$  (\ref{5}) and
 $G^{\tau\tau}={1}/{ L^{2}(1-\dot{\vec{x}}^2)}={1}/{ L^{2}G^{tt}}$ 
 we express $\mathcal{P}_0$ as a function of the velocity $\dot{\vec{x}}$
\begin{equation}\label{5'}
\mathcal{P}_0:=TL\sqrt{|detG|}G^{\tau\tau}=T\sqrt{\frac{|g|}{1-\dot{\vec{x}}^2}} \,\,\,.
\end{equation}
Taking into account this expression 
and definition (\ref{5}) one 
can present $\vec{\mathcal{P}}$ and its evolution equation (\ref{5}) in the form 
previously used in \cite{Hops}, \cite{TZ} and \cite{Znpb},
\begin{equation}\label{13}
\vec{\mathcal{P}}=\mathcal{P}_0 \dot{\vec{x}}
 , \ \ \ \
\dot{\vec{\mathcal{P}}}= T^{2}\partial_r \left( \frac{|g|}{\mathcal{P}_0} g^{rs}\partial_s 
\vec{x}\right).
\end{equation}
Then Eqs. (\ref{13}) produce the second-order PDE for $\vec{x}$
\begin{equation}\label{xeqv}
 \ddot{\vec{x}}
=\frac{T}{\mathcal{P}_0}\partial_r \left( \frac{T}{\mathcal{P}_0}|g|g^{rs}
\partial_s \vec{x}\right).
\end{equation}
These equations may be presented in the canonical Hamiltonian  form
\[
\dot{\vec{x}}=\{H_{0}
,\vec{x}\}, \ \ \ \ \dot{\vec{\mathcal{P}}}=\{H_{0},\vec{\mathcal{P}}\}, \ \ \ \
\{\mathcal{P}_i(\sigma), x_j(\tilde{\sigma})  \}=
\delta_{ij}\delta^{(p)}(\sigma^r-\tilde{\sigma}^r),
\]
where $H_{0}$ is the integrated Hamiltonian density $\mathcal{H}_0\equiv\mathcal{P}_0$ 
\begin{eqnarray}\label{hampbr}
H_{0}=  \int d^p \sigma \sqrt{\vec{\mathcal{P}}^2+T^{2}|g|}.
\end{eqnarray}
The presence of square root in (\ref{hampbr}) points to the presence of the known 
residual symmetry preserving the orthogonal gauge (\ref{7}) 
\begin{equation}\label{diff}
\tilde{t}=t, \ \ \ \ \tilde{\sigma}^r=f^r(\sigma^s)
\end{equation}
and generated by the  constraints $\tilde{T}_r$  (\ref{6}) reduced to the form
\begin{equation}\label{T}
T_r:=\vec{\mathcal{P}}\partial_r\vec{x}=0 \ \ \    \Leftrightarrow \ \ \ 
 \dot{\vec{x}} \partial_r\vec{x}=0,\ \ \ (r=1,2, \ldots, p).
\end{equation}
 The freedom allows to impose
 $p$ additional time-independent conditions on $\vec{x}$ and 
 its space-like derivatives. The presented description does not restrict 
 space-time and brane worldvolume dimensions $(D,p)$ and  $p<D$. 

Alternatively, we present $p$-brane Eqs. (\ref{5}) 
as the reparametrization invariant wave equation for $x^{m}$ on 
the $(p+1)$-dim. brane worldvolume $\Sigma_{p+1}$
\begin{equation}\label{Box}
\Box^{(p+1)}x^{m}=0, 
\end{equation}
where  $\Box^{(p+1)}:
=\frac{1}{\sqrt{|G|}}\partial_{\alpha} \sqrt{|G|}G^{\alpha\beta}\partial_{\beta} 
$ is the Laplace-Beltrami operator. 

Using the relation 
$
\partial_{\alpha}\ln\sqrt{|G|}= \Gamma_{\alpha\beta}^{\beta},
$
where $\Gamma_{\alpha\beta}^{\gamma}$ are the Cristoffel symbols 
generated by the metric  $G_{\alpha\beta}$ of $\Sigma_{p+1}$,
 one can express Eqs.(\ref{Box}) as the vanishing 
covariant divergence of the worldvolume vector $x^{m,\alpha}$
\begin{equation}\label{covarder}
\Box^{(p+1)}x^{m}\equiv \nabla_{\alpha} x^{m,\alpha}=0,
\end{equation}
where  $ x^{m,\alpha}:= G^{\alpha\beta}\partial_{\beta}x^{m}$
 and
$ \nabla_{\alpha}x^{m,\alpha}\equiv \partial_{\alpha} x^{m,\alpha} + 
\Gamma_{\beta\alpha}^{\alpha} x^{m,\beta}$.
Eqs.(\ref{covarder}) are presented as the continuity equations 
 $$\partial_{\alpha}T^{m\alpha}=0$$
for the components of Noether current 
$T^{m\alpha}:=T\sqrt{|G|}G^{\alpha\beta}\partial_{\beta}x^{m}$ 
generated by the global translation symmetry of the  Minkowski target space.

Below, we shall use the wave representation (\ref{Box}) 
in a fixed gauge to develop 
a way for construction of some exact solutions of the brane equations.

\section {Laplace-Beltrami operator and \\ 
Noether identities for p-branes}

Using the gauge (\ref{7}) one can extract the
 Laplace-Beltrami operator $\Delta^{(p)}$, associated with  the p-brane 
hypersurface $\Sigma_{p}$, from the operator $\Box^{(p+1)}$
\begin{equation}\label{LaB}
\Delta^{(p)}\vec{x}:=
\frac{1}{\sqrt{|g|}}\partial_{r}\left( \sqrt{|g|}g^{rs}\partial_s \vec{x} \right), 
\end{equation}
where $g_{rs}:=\partial_r \vec{x} \cdot \partial_s \vec{x}$ is the induced metric 
on $\Sigma_{p}$. The use of the  LB operator $\Delta^{(p)}$ 
 allows to present Eqs. (\ref{Box}) as the system of $(D-1)$ equations 
\begin{equation}\label{rxeqv}
\ddot{\vec{x}}=\frac{1}{2}\partial_r(1 - {\dot{\vec{x}}}^2)
 g^{rs} \partial_s\vec{x}
+(1 - {\dot{\vec{x}}}^2)\Delta^{(p)}\vec{x}.
\end{equation}
Taking into account the relation 
$
\frac{1}{2}\partial_r(1 - {\dot{\vec{x}}}^2)=(\ddot{\vec{x}}\partial_r\vec{x}),
$
following from the orthogonality conditions (\ref{T}), we rewrite the system 
(\ref{rxeqv}) in the form  
\begin{equation}\label{xeqvm}
\ddot{\vec{x}} - (\ddot{\vec{x}}\vec{x}^{,r})\vec{x}_{,r}
= (1 - {\dot{\vec{x}}}^2)\Delta^{(p)}\vec{x},
\end{equation}
where the following condensed notations are used
\begin{equation}\label{sogl}
\vec{x}_{,r}:=\partial_{r}\vec{x}, \ \ \
\vec{x}^{,r}:=g^{rs}\vec{x}_{,s} \ \  \rightarrow  \ \  \vec{x}_{,r}\vec{x}^{,s}=\delta_{r}^{s}.
\end{equation}
Eqs. (\ref{xeqvm}) show equality between two invariants 
of the residual diffeomorphisms (\ref{diff}) of $\Sigma_p$ 
one of which is $\Delta^{(p)}\vec{x}$, including only the space-like 
derivatives of $\vec{x}$, and the other 
\begin{equation}\label{inv2}
I:= G^{tt}[ \ddot{\vec{x}} - (\ddot{\vec{x}}\vec{x}^{,r})\vec{x}_{,r}]
\end{equation}
 capturing all time-like derivatives of $\vec{x}$.
$I$ equals the metric component $G^{tt}= 1/(1 - {\dot{\vec{x}}}^2)$ multiplied by 
the l.h.s. of Eqs.(\ref{xeqvm}) equal to projection of the  
acceleration  $\ddot{\vec{x}}$ on the directions orthogonal 
to $\Sigma_{p}$. This follows from the identities
\begin{equation}\label{prolap}
\vec{x}_{,r}[\ddot{\vec{x}} - (\ddot{\vec{x}}\vec{x}^{,s})\vec{x}_{,s}]\equiv0
\end{equation}
that imply that $(\vec{x}_{,r}\Delta^{(p)}\vec{x})\equiv0$  
which are a consequence  of the formula  
\begin{equation}\label{lapexp}
\Delta^{(p)}\vec{x}\equiv\nabla_{s} \vec{x}^{,s}=\partial_s\vec{x}^{,s} +
(\partial_sln\sqrt{|g|})\vec{x}^{,s}.
\end{equation}
 The covariant derivative
 $\nabla_{r} \vec{x}^{,s}:=\partial_{r} \vec{x}^{,s} + \Gamma_{rq}^{s} \vec{x}^{,q}$ 
 contains the Cristoffel symbols $\Gamma_{ps}^{r} $ constructed 
 from the metric tensor $g_{rs}$ of the brane hypersurface $\Sigma_{p}$.
 Indeed, the representation (\ref{lapexp}) multiplied by
  the vectors $\vec{x}_{,r}$ results in 
\begin{equation}\label{prolap'}
(\vec{x}_{,r}\Delta^{(p)}\vec{x})=\partial_{r}ln\sqrt{|g|} + 
\vec{x}_{,r}\partial_s\vec{x}^{,s}= \partial_{r}ln\sqrt{|g|} 
- \frac {1}{2}(g^{sq}\partial_{r}g_{qs})\equiv0
\end{equation}
in view of the well-known relation $g^{sq}dg_{qs}=dln|g|$.
The derived identities (\ref{prolap}) extracted from Eqs.(\ref{xeqvm})
 are the Noether identities associated with the residual gauge 
 symmetry (\ref{diff}) of the $p$-brane equations. 
 
 From the physical point of view the brane Eqs.(\ref{xeqvm})
 mean that the constituent of $\ddot{\vec{x}}$ 
 orthogonal to $\Sigma_p$ is parallel to $\Delta^{(p)}\vec{x}$, 
 and therefore the forces orthogonal to the brane hypersurface 
 are represented by the vector $\Delta^{(p)}\vec{x}$.
 The geometric interpretation of the invariant $I$ allows to express 
 the brane equations (\ref{xeqvm}) in the equivalent form
\begin{equation}\label{xeqmp}
\Pi_{ik}\ddot x_{k}=(1 - {\dot{\vec{x}}}^2)\Delta^{(p)}x_{k},
\end{equation}
where $\Pi_{ik}$ is the projection operator 
\begin{equation}\label{proj}
\Pi_{ik}:=\delta_{ik}- x_{i,r}x_{k}^{,r}, \ \ \ \Pi_{ik}\Pi_{kl}=\Pi_{il}
\end{equation}
on the local vectors $\vec{n}_{\perp}$ 
orthogonal to the tangent vectors $\vec{x}_{,r}$ of $\Sigma_p$. 
Then the property of orthogonality of $\Delta^{(p)}\vec{x}$  to $\Sigma_p$ is encoded by the conditions 
\begin{equation}\label{lappro}
\Pi_{ik}\Delta^{(p)}x_{k}=\Delta^{(p)}x_{i}
\end{equation}
showing that $\Delta^{(p)}\vec{x}$ is an 
eigenvector of the projection operator $\Pi_{ik}$ 
similarly to  the Euclidean vectors $\vec{n}_{\perp}$  and  $\dot{\vec{x}}$
\begin{equation}\label{velpro}
 \Pi_{ik}\dot x_{k}=\dot x_{i},  \ \ \   \Pi_{ik}n_{\perp k}=n_{\perp i}.
\end{equation}

The presence of $p$ Noether identities (\ref{prolap}) proves that 
 $(D-1)$  brane equations (\ref{xeqvm}) contain only $(D-p-1)$ independent equations  
\begin{equation}\label{prorg}
\vec{n}_{\perp}[ \ddot{\vec{x}} - (1- \dot{\vec{x}}^2) g^{rs}\vec{x}_{,rs}] =0,
\end{equation}
generated by the projections of (\ref{xeqvm}) on the 
vectors $\vec{n}_\perp(t,\sigma^{r})$ orthogonal 
to the tangent  hyperplane spanned by the vectors $\partial_{r}\vec{x}$ 
 at the point $(t,\sigma^{r})$
\begin{equation}\label{newrel}
\vec{n}_{\perp}\vec{x}_{,s}=0 \ \ \rightarrow \ \ 
\vec{n}_{\perp}\partial_{r}\vec{x}^{,r}=g^{rs}(\vec{n}_{\perp}\vec{x}_{,rs}),
\end{equation}
where the subindex $\perp=p+1,p+3,..., D-1$ takes $(D-p-1)$ values.

Using $G_{\alpha\beta}$ (\ref{7}) one can  present 
Eqs.(\ref{prorg}) in an equivalent form 
\begin{equation}\label{minim}
G^{\alpha\beta}W^{\perp}_{\alpha\beta}
\equiv G^{\alpha\beta}(\vec{n}_{\perp}\vec{x}_{,\alpha\beta})=0
\end{equation}
recognized as  the minimality conditions for the
worldvolume $\Sigma_{p+1}$ embedded in the $D$-dimensional Minkowski 
space expressed via the covariant traces of the second fundamental 
form $W^{\perp}_{\alpha\beta}$ of the brane worldvolume $\Sigma_{p+1}$.

In the considered orthogonal gauge (\ref{7}) the $(p+1)$-st Noether identity,  
associated with the freedom in $\tau$-reparametrizations 
 of $\Sigma_{p+1}$, reduces to the energy density 
conservation $\dot{\mathcal{P}}_0=0$. It can be seen when analyzing the projection 
 of (\ref{xeqvm}) on the vector $\dot{\vec{x}}$. 
Really, taking into account the relations
\begin{eqnarray}\label{prolapt}
\dot{\vec{x}} \Delta^{(p)}\vec{x}=-\frac {1}{2}(g^{sq}\partial_{t}g_{qs})
=-\partial_{t}ln\sqrt{|g|},\\
\dot{\vec{x}}\,\frac{\ddot{\vec{x}} 
- (\ddot{\vec{x}}\vec{x}^{,s})\vec{x}_{,s}}{1- \dot{\vec{x}}^2}
=-\partial_{t}ln\sqrt{1- \dot{\vec{x}}^2}. \label{prolapt´} 
\end{eqnarray}
one can present the projection of Eqs.(\ref{xeqvm}) on $\dot{\vec{x}}$ as
\begin{equation}\label{protim} 
\partial_{t}ln\sqrt{1- \dot{\vec{x}}^2}= \partial_{t}ln\sqrt{|g|}
\end{equation}
or, after using definition (\ref{5'}) for the energy density $\mathcal{P}_0$, in the form 
\begin{equation}\label{protim'} 
\partial_{t}ln\sqrt{\frac{|g|}{1-\dot{\vec{x}}^2}}=\partial_{t}ln(\frac{\mathcal{P}_0}{T})=0.
\end{equation}  
Eq. (\ref{protim'}) is satisfied  in view of the above-proved energy conservation law.

\section {Solvable $p$-brane motions with $\Delta^{(p)}\vec{x}=0$}

The interpretation of $\Delta^{(p)}\vec{x}$ as the vector encoding
 forces orthogonal to $\Sigma_p$  may be used for exploring admissible 
 motions of branes. On this way it is natural to study the  motions 
 in the absence of forces orthogonal to the brane hypersurface $\Sigma_p$. 
 These motions are fixed by the harmonicity conditions
 \begin{equation}\label{statn}
 \Delta^{(p)}\vec{x}=0
 \end{equation}
 which must be considered as the initial value constraints 
 for brane Eqs.(\ref{xeqvm}).
 Since the  constraints (\ref{statn}) have to be preserved in time 
 the corresponding brane evolution must obey 
the following equations  
 \begin{equation}\label{timevol} 
\ddot{\vec{x}} - (\ddot{\vec{x}}\vec{x}^{,r})\vec{x}_{,r}=0,
\end{equation}
as it follows from Eqs.(\ref{lappro}). 
It is easy to see that Eqs. (\ref{timevol}) have a particular solution  
that coincides with the general solution of the system
\begin{equation}\label{trans}
\ddot{\vec{x}}=0, \ \ \ \ \  \Delta^{(p)}\vec{x}=0
\end{equation}
which describes the motions in the balance of forces acting on the brane.
The general solution of evolution Eqs. (\ref{trans}) is linear in time 
\begin{equation}\label{soltrans}
\vec{x}= \vec{x}_{0}(\sigma^r) + \vec{v}_{0}(\sigma^r)t,  \ \ \
 \vec{v}_{0}\vec{x}_{0,r}=0,   \ \ \  \vec{v}_{0}^2= constant,
\end{equation}
as it follows from the orthogonality conditions (\ref{T}).
Then harmonicity conditions (\ref{trans}) are transformed to 
constraints for the initial values $\vec{x}_{0}(\sigma^r)$ 
and $\vec{v}_{0}(\sigma^r)$.  
The static $p$-branes are described by the particular solution 
\begin{equation}\label{stat}
\vec{x}=\vec{x}_{0}(\sigma^r),  \ \ \ \ \  \Delta^{(p)}\vec{x}_{0}=0 
\end{equation}
and the harmonicity conditions yield the initial data constraints 
for the brane shape $\vec{x}_{0}(\sigma^r)$.
The static brane energy density  $\mathcal{P}^{(stat)}_{0}=T\sqrt{|g|}$ 
and it can realize the ground state of $p$-brane, as its kinetic energy vanishes. 
Let us note that an antipode of the static brane is the one
moving with the maximum velocity equals the velocity of light, i.e. 
$\dot{\vec{x}}^2=1$. In this case Eqs. (\ref{xeqvm}) are reduced to 
the above-discussed equation $\ddot{\vec{x}}=0$, but with 
arbitrary  $\Delta^{(p)}\vec{x}$.
The branes moving with the velocity of light have  zero tension  
and degenerate  metric (\ref{7}) of their worldvolumes \cite{Z_0}. 
The discussed  examples of particular solutions confirm 
correctness of the proposed approach for exploring solutions of 
Eqs. (\ref{xeqmp}). So, one can apply it for studying the general 
solution of (\ref{timevol}) describing 
 tensionfull branes characterized by $(1- \dot{\vec{x}}^2)>0$. 

Generally Eqs.(\ref{timevol}) capture the whole set 
of motions characterized by zero projections of the 
 acceleration $\ddot{\vec{x}}$ on the directions orthogonal to $\Sigma_{p}$ 
\begin{equation}\label{lap0}
\Delta^{(p)}\vec{x}=0=\ddot{\vec{x}} - 
(\ddot{\vec{x}}\vec{x}^{,r})\vec{x}_{,r} \ \ \
 \longrightarrow \ \ \  \ddot{\vec{x}} \dot{\vec{x}}=0 \ \ \  
\longrightarrow \ \ \  \dot{\vec{x}}^2=\vec{v}^2(\sigma^r).
\end{equation}
The forces acting on the brane are tangent 
to  $\Sigma_{p}$ and produce acceleration
 orthogonal to the velocity $\dot{\vec{x}}$, respectively.
Combining the time-independence of both the squared velocity $\dot{\vec{x}}^2$
 and the energy density we obtain the formula
\begin{equation}\label{gensden'}
\mathcal{P}_0(\sigma^r)=T\sqrt{\frac{|g|}{1-\vec{v}^2(\sigma^r)}}.
\end{equation}
which shows time-independence of the brane volume, i.e.  $\dot{g}=0$. 
These conditions are characteristic of spinning $p$-branes with
 their elastic force compensated by the centrifugal force. 
This proves that the solutions of the equations $\Delta^{(p)}\vec{x}=0$ 
must describe spinning $p$-branes.
  To find such solutions in explicit form we restrict ourselves 
 by the case when spinning  $p$-branes evolve in odd-dimensional 
Minkowski space with the fixed dimension $D=(2p+1)$. 

In this case we have $p$ independent components of $\vec{x}(t,\sigma^{r})$ 
remaining after the solution of the $p$ orthogonality 
constraints $(\dot{\vec{x}} \cdot \partial_r \vec{x})=0$.
In view of the above-derived $p$ Noether identities we have just $p (=2p-p)$ 
independent equations for $p$ remaining degrees of freedom of
$\vec{x}(t,\sigma^{r})$. In addition there are $p$  $\sigma$-dependent 
diffeomorphisms (\ref{diff}) which can be used to fix $\sigma$-dependence 
of these DOF. Finally, the brane equations are reduced to the system of $p$ 
usual differential equations for $p$ functions independent of $\sigma^r$.
A possible way to accomplish such a type of reduction is, e.g. to separate $t$ 
and $\sigma$ variables in each component of the vector $\vec{x}(t,\sigma^{r})$ 
\begin{equation}\label{gensden}
x_{i}(t,\sigma^{r})=u_{i}(t)v_{i}(\sigma^{r})
\end{equation}
with subsequent exclusion of gauge and non-propagating DOF using $p$ 
orthogonality conditions (\ref{7}) and $p$ additional gauge conditions 
for the remaining diffeomorphisms (\ref{diff}).
  This strategy  was realized in \cite{IJGM}, where the 
discussed $2p$-dimensional Euclidean  vector $\vec{x}(t,\sigma^{r})$ 
 of spinning ${p}$-brane was presented as the generalization of
 the membrane anzatses studied in \cite{JU1} and \cite{TZ} 
\begin{eqnarray} 
\vec{x}^T(t,\sigma^{r})
=(q_1\cos\theta_1,q_1\sin\theta_1,q_2\cos\theta_2,q_2\sin\theta_2,
\ldots,q_p\cos\theta_p,q_p\sin\theta_p), \label{spinanzats}
 \\
q_a=q_a(\sigma^{r}),\,\,\, 
 \theta_a=\theta_a(t)
\, \, \, \, \,  \, \, \,\, \, \, \, \, \, \, \,\, \, \, \, \, \, \, \, \, \, \, \, \, 
\, \, \, \, \, \, \, \, \, \, \, \, \, \, \, \,  \, \, \,\nonumber
\end{eqnarray}
which gives a solution of orthogonality constraints (\ref{7})  with 
the  propagating DOFs  represented by the polar angles  $\theta_{a}(t)$. 
This anzats gives 
$$\dot{\vec{x}}^2= \sum_{a=1}^{p}q_{a}^{2}\dot\theta_{a}^{2}.$$
Keeping in mind constraint (\ref{lap0}) we obtain the following solution for $\theta_{a}(t)$
 \begin{equation}\label{solteta}
 \sum_{a=1}^{p}q_{a}^{2}\dot\theta_{a}^{2}(t)=\vec{v}^2(\sigma^r) \longrightarrow \ \ \ 
 \theta_{a}(t) =\theta_{a0} + \omega_{a}t, \ \ \  
\end{equation}
where $\theta_{0a}$ and  $\omega_{a}$  are the integration 
constants with a=1,2,...,p. 

As a result, the energy density of spinning $p$-brane
 $\mathcal{P}_0$ (\ref{gensden}) 
is defined by the  following function of the velocity 
components $\omega_{a}q_{a}(\sigma^{r})$
\begin{equation}\label{5spin}
\mathcal{P}_0=T\sqrt{\frac{|g|}{1-\sum_{a=1}^{p}\omega_{a}^{2}q_{a}^{2}}}.
\end{equation}
This time-independent energy density turns into the  density  $\mathcal{P}^{(stat)}_{0}$ 
of a static brane 
in the limiting case of all the vanishing frequencies: $\omega_{a}=0$.

The separation between $t$ and $\sigma^{r}$ variables realized  by
 anzats (\ref{spinanzats}) turns out to be a sufficient  condition 
for exact solvability of Eqs.(\ref{spinanzats}).
Indeed, the substitution of  (\ref{spinanzats}) into (\ref{xeqvm})
 reduces these $2p$ nonlinear PDEs for the components of $\vec x$
to $p$ PDEs for the $p$ components of $\mathbf{q}(\sigma^{r}):=(q_1,..,q_p)$.
\begin{eqnarray}\label{spineqv}
-\omega_{a}^2q_{a} + \sum_{b,r,s=1}^{p}\omega_{b}^2q_{b}(q_{b,r}g^{rs}q_{a,s})=
(1- \sum_{b=1}^{p}q_{b}^2\omega_{b}^2)\Delta^{(p)}q_{a},
\end{eqnarray}  
Because $\Delta^{(p)}q_{a}=0$, as a consequence of $\Delta^{(p)}x_{m}=0$,  
Eqs. (\ref{spineqv}) are satisfied if there is exact cancellation 
between all its terms. The cancellation occurs when the conditions 
\begin{equation}\label{pbein}
g^{rs}q_{a,r}q_{b,s}=\delta_{ab}, \ \ \ \   
g_{rs}=q_{a,r}q_{a,s}
\end{equation} 
for the induced metric $g_{rs}$  on $\Sigma_{p}$ generated 
by (\ref{spinanzats}) are satisfied.
 These conditions  
express  the space-like part of metric (\ref{7}) exactly in the form
 connecting its with the components of the  $p$-bein $e^{a}_{r}$ attached 
to the  hypersurface $\Sigma_{p}$. 

As a result, the partial derivatives $q^{a}_{,r}$  
  coincide with  the  $p$-bein $e^{a}_{r}$ and conditions (\ref{pbein})
 may be presented in the equivalent form as
\begin{equation}\label{pbein'}
e^{a}_{r}=q^{a}_{,r}. 
\end{equation} 
The worldvolume metric $G_{\alpha \beta}$ on $\Sigma_{p+1}$ generated 
by anzats (\ref{spinanzats}) is given by 
\begin{eqnarray}\label{metrik} 
 G_{tt}=1- \sum_{a=1}^{p}q_{a}^2\omega_{a}^2, \  \  \  
g_{rs}= \sum_{a=1}^{p}q_{a,r}q_{a,s}\equiv \mathbf{q}_{,r}\mathbf{q}_{,s},  
\  \  \  
\mathbf{q}:=(q_1,..,q_p)
\end{eqnarray}  
 which  yields the following  squared 
interval $ds^2_{p+1}$ on $\Sigma_{p+1}$
\begin{equation}\label{intr}   
ds^2_{p+1}=
(1- \sum_{a=1}^{p}q_{a}^2\omega_{a}^2) dt^2
 - \sum_{a=1}^{p}dq_{a}dq_{a}.
\end{equation}
 This shows that in terms of the new coordinates $q_{a}(\sigma^{r})$ 
the hypersurface $\Sigma_{p}$ metric $g_{rs}$ becomes independent 
of $\sigma^{r}$.

For infinite $p$-branes without boundary conditions 
and $-\infty<\sigma^{r}<+\infty$ one can choose the 
following gauge for the residual symmetry (\ref{diff})
\begin{equation}\label{hpgaug}
q_{1}(\sigma^{r})=k\sigma^{1},\ \ \ 
 q_{2}(\sigma^{r})=k \sigma^{2},\ ....\ ,
q_{p}(\sigma^{r})=k\sigma^{p},
\end{equation}
where $k\sim T^{\frac{-1}{p+1}}$ is an arbitrary constant with 
the dimension of length. 
This choice results in the constant diagonal 
matrices for $p$-bein $e^{a}_{r}$ and metric $g_{r,s}$
\begin{equation}\label{flat}
e^{a}_{r}=k\delta^{a}_{r},\ \ \ \ g_{rs}=k^2\delta_{rs}
\end{equation}
which solve the considered harmonic 
equations $\Delta^{(p)}q_{a}(\sigma^r)=\Delta^{(p)}\vec{x}(t,\sigma^r)=0$.

It proves that the initial value constraints $\Delta^{(p)}\vec{x}=0$ 
select exact solutions of Eqs.(\ref{xeqmp}) describing spinning  
branes with the shape of $p$-dim. hyperplanes
\begin{eqnarray}\label{inf} 
\vec{x}^T(t, \sigma^{r})
=k(\sigma^{1}\cos(\theta_{10}+\omega_{1}t), \, \sigma^{1}\sin(\theta_{10}
+\omega_{1}t),
\ldots,  \\    \nonumber
\sigma^{p}\cos(\theta_{p0}+\omega_{p}t), \, \sigma^{p}\sin(\theta_{p0}
+\omega_{p}t) )
\end{eqnarray}
The energy density of the infinite spinning branes is given by 
\begin{equation}\label{5spin´}
\mathcal{P}_0(\sigma^r)=\frac{Tk^p}
{\sqrt{1-k^2\sum_{a=1}^{p}\sigma_{a}^{2}\omega_{a}^{2}}} 
\end{equation}
and one can see that the 
condition $k\omega\sim \omega T^{\frac{-1}{p+1}}\rightarrow 0$
has to be satisfied when $|\sigma_{a}|\rightarrow \infty$ to 
preserve the real value of $\mathcal{P}_0$.
 This demands $\omega\rightarrow 0$ when the tension $T$ is fixed, 
and thus $\mathcal{P}_0$ (\ref{inf}) becomes a 
constant $\sim T^{\frac{1}{p+1}}$ resulting in the divergent total 
energy in the static limit because of the infinite  integration range 
 in the parameters $\sigma_{a}$. The static solutions may be treated as
 domain "hyperwalls" generalizing the well-known two-dimensional domain 
walls which appear as solutions in various physical models.

The integration range in $\sigma$ can be made a compact by 
considering closed or open branes with the corresponding boundary 
conditons. Below we consider the case of closed spinning $p$-branes 
described by the anzats (\ref{spinanzats}).

\section {Folded  $p$-branes as solutions of $\Delta^{(p)}\vec{x}=0$}

The change of gauge conditions (\ref{hpgaug}) into the ones 
considered in \cite{IJGM}
\begin{equation}\label{rgaug}
q_{1}(\sigma^{r})=q_{1}(\sigma^{1}),\ \ \  q_{2}(\sigma^{r})=
q_{2}(\sigma^{2}),\ ....\ ,\ q_{p}(\sigma^{r})=q_{p}(\sigma^{p}), 
\end{equation}
where each of the functions $q_{a}$ is a monotonic continuous function 
of only the variable $\sigma^r$ with $r=a$, gives more general 
solutions for conditions (\ref{pbein})
\begin{eqnarray}
q_{a,r}=\delta_{as}\acute{q}_{r},\ \ \  \acute{q}_{s}:=\frac{dq_s}{d\sigma^s}, 
  \ \ \ \ \ \ \ \ \ \ \ \ \ \ \ \ \ \
\nonumber  \\
g_{rs}=\delta_{rs}\acute{q}_{s}^{2}, \ \ 
 g^{rs}=\frac{\delta_{rs}}{\acute{q}_{s}^{2}},\ \ \
g= \prod_{a=1}^{p}\acute{q}_{a}^{2}\equiv \prod\acute{q}_{a}^{2} \label{metrpbe}
\end{eqnarray}
with the diagonal matrices $q_{a,r}$ and  $g_{rs}$, and factorized 
determinant of $g_{rs}$.
The radial components $q_{a}(\sigma^r)$ (\ref{rgaug})
and metric (\ref{metrpbe}) are the solutions of  eqs. $\Delta^{(p)}\vec{x}=0$. 
To verify the statement it is enough to prove that these $q$-coordinates 
are the solutions of the reduced harmonic equations 
\begin{equation}\label{zerlapl}
\Delta^{(p)}q_{a}(\sigma^r)=0,   \ \ \ \  (a=1,2,...,p).
\end{equation}
This becomes  evident after the substitution of (\ref{metrpbe}) 
into (\ref{zerlapl}) resulting in 
\begin{equation}\label{LaBq'}
\Delta^{(p)}q_{a}=
\frac{1}{\prod\acute{q}_{b}}
\frac{\partial}{\partial\sigma^{a}}
\left( 
\frac{\prod\acute{q}_{b}}{\acute{q}_{a}} 
\right)=0.
\end{equation}
 The latter equations are satisfied in view of cancellation of the 
derivative  $\acute{q}_{a}$ which is only one 
function depending on $\sigma^a$ in the 
fraction  $\frac{\prod\acute{q}_{b}}{\acute{q}_{a}}$. 

It is clear, that the mapping (\ref{rgaug}) with regular 
 monotonic $q$-functions describes the same 
infinite $p$-dimensional hyperplanes as the solution (\ref{hpgaug}).

However, the replacement of the monotonic $q$-functions by the
 periodic ones with isolated nonregular points  in  $g^{rs}$ 
(\ref{metrpbe}) yields solutions of Eqs. $\Delta^{(p)}\vec{x}=0$ 
describing compact folded $p$-branes. 
The solutions generalize ones describing the folded strings \cite{ZYaF},
 \cite{Hops}, \cite{NNK} to the case of $p$-branes. 
The folds arise as a result of the one-parametric
 dependence of the functions $q_{a}(\sigma^a)$ (\ref{rgaug})
 applied to describe closed $p$-brane by the generalized 
anzats (\ref{inf})
\begin{eqnarray}\label{ganzats} 
\vec{x}^T(t, \sigma^{r})
=(q_1(\sigma^{1})\cos(\theta_{10}+\omega_{1}t), \, q_1(\sigma^{1})\sin(\theta_{10}
+\omega_{1}t),
\ldots,  \\    \nonumber
q_p(\sigma^{p})\cos(\theta_{p0}+\omega_{p}t), \, q_p(\sigma^{p})\sin(\theta_{p0}
+\omega_{p}t) )
\end{eqnarray}
 with the initial data $\theta_{0a}=0$ at $t=0$ and the density 
 energy (\ref{5spin}) given by 
\begin{equation}\label{5spinen}
\mathcal{P}_0=\frac{T|\prod\acute{q}_{a}|}
{\sqrt{1-\sum_{a=1}^{p}q_{a}^{2}\omega_{a}^{2}}} \ .
\end{equation}
In the case of closed $p$-branes their  $\sigma$-parameters 
are bounded:  $\sigma^r\in [0,2\pi]$, and therefore each of the 
 functions $q_{a}(\sigma^r)$ from (\ref{ganzats}) has to be 
a periodic one:  $q_{a}(0)=q_{a}(2\pi)$. 
Next we see that at any moment $t$  the 
 world vector $\vec{x}^T(t,\sigma^r)$ (\ref{ganzats}) 
is produced from $\vec{x_0}^T(\sigma^r)=(q_1,0,q_2,0,\ldots,q_p,0 )$ 
by the time-parametrized rotations belonging to
 the diagonal subgroup $U(1)^p$ of the group $SO(2p)$.
This subgroup is composed of the time-dependent 
rotations in the planes $x_1x_2$, $x_3x_4$ ,...,$x_{2p-1}x_{2p}$
about the angles $\theta_{a}=\theta_{0a}+\omega_{a}t$, respectively.
Thus, the $p$-brane worldvolume is formed by the rotations 
of the closed $p$-brane initially embedded into the $p$-dim. 
subspace spanned by all odd coordinate axises of the 
considered $2p$-dim. Euclidean space. These rotations preserve 
the initial brane shape. So, the periodicity conditions  
for $q_1$ with respect to $\sigma_1$, $q_2$ with respect to $\sigma_2$,
 etc. will be satisfied if the $p$-brane is initially  folded up along 
each of the odd coordinate axises. 
A simple example of the solution is given 
 by the symmetrically folded closed $p$-brane 
\begin{eqnarray}\label{folbr} 
\vec{x}^T(0,\sigma^r)
=k(|\pi-\sigma^{1}|, 0, 
|\pi-\sigma^{2}|, 0,|\pi-\sigma^{2}|,\ldots, |\pi-\sigma^{p}|, 0)
\end{eqnarray}
with the functions $q_{a}(\sigma^{a})=k|\pi-\sigma^{a}|$ 
which realize the conditions $q_{a}(0)=q_{a}(2\pi)$ by the 
bending formation at $\sigma^{a}=\pi$ which create additional forces 
orthogonal to $\Sigma_p$ around these points. The latters fix   
the lines (planes) on the brane hypersurface $\Sigma_p$ 
along which it is bent. 
For the folded membrane ($p=2$) embedded into 4-dim. 
Euclidean space its image may be visualized as a double-folded sheet  
of paper forming a stack of four equal small squares originated  
from the original unfolded square with the side length equal to $2k\pi$.
The functions  $q_{a}(\sigma^{a})$  in (\ref{folbr}) are continuous ones, 
but their derivatives have the jump discontinuity 
equal to $2=1-(-1)$ at  $\sigma^{a}=\pi$. 
These jumps result in the indefiniteness of the induced 
metric (\ref{metrpbe}) at these points. 
The change of the parametrization (\ref{folbr}) by
\begin{eqnarray}\label{folbr1} 
\vec{x}^T(0,\sigma^r)
=k(\sin\frac{\sigma^{1}}{2}, 0, 
\sin\frac{\sigma^{2}}{2}, 0, \ldots, \sin\frac{\sigma^{p}}{2}, 0)
\end{eqnarray}
 smooths out the derivative jumps at $\sigma^{a}=\pi$.
  The flat metric $g_{rs}$ (\ref{metrpbe})  vanish 
at these points, as well as the energy 
density $\mathcal{P}_0$ (\ref{5spinen}) 
(if $\sum_{a=1}^{p}q_{a}^{2}\omega_{a}^{2}\neq 1$).

A more general parametrization producing $(n_{1},n_{2}, \ldots, n_{p})$ 
singular points for  $g^{rs}$ defined by the 
functions $(q_{1},q_{2}, \ldots, q_{p})$ (\ref{metrpbe}), 
respectively, may be choosen in the form similar to the one considered 
in \cite{ZYaF}
\begin{eqnarray}\label{folbrn} 
\vec{x}^T(0,\sigma^r)
=k(\sin\frac{n_1\sigma^{1}}{2}, 0, 
\sin\frac{n_2\sigma^{2}}{2}, 0, \ldots, \sin\frac{n_p\sigma^{p}}{2}, 0)
\end{eqnarray}
with the set $(n_{1},n_{2}, \ldots, n_{p})$ treated as the topological 
winding numbers.

So, anzats  (\ref{ganzats}) with the periodic $q$-functions gives exact
 solutions of $\Delta^{(p)}\vec{x}=0$ with isolated singularities 
in $g^{rs}$ and describe initially folded branes.
The brane worldvolume $\Sigma_{p+1}$ associated  with the initially   
folded hypersurface $\Sigma_p$ is produced by its 
rotations as a whole realized by the above mentioned Abelian group 
$U(1)\times U(1)\times\ldots \times U(1)\equiv U(1)^p$. 
The corresponding rotation angles  $\theta_{a}$ 
are treated as the generalized cyclic coordinates of the Hamiltonian 
density (\ref{9})
corresponding to the energy density $\mathcal{P}_0$ (\ref{5spin}).
The momenta $j_a$ conjugate to the generalized coordinates $\theta_a$
$$
j_a:=\frac{\partial \mathcal{L}}{\partial \dot{\theta_a}}=
\vec{\mathcal{P}}\frac{\partial{\dot{\vec{x}}}}{\partial \dot{\theta_a}}
$$
  are given by 
\begin{equation}\label{tetimp}  
 j_a=\mathcal{P}_0 q_{a}^2\dot{\theta_a}
\equiv\mathcal{P}_0 \omega_{a}q_{a}^2 .
\end{equation}
Then the corresponding Hamiltonian $p$-brane density takes the form
\begin{equation}\label{rotahamd}
\mathcal{H}_0=\sqrt{\sum_{a=1}^{p}(j_{a}/q_{a})^2 + T^{2}|g|} \ . 
\end{equation}
The momenta (\ref{tetimp}) are integrals of the motion
$$
\frac{d j_a}{dt}=0,  \ \ \ \ \ \   (a=1,2,..,p)
$$
 proportional to the conserved energy density  $\mathcal{P}_0$.
The values  $j_a$ are the components of the angular momentum 
density associated with the generators of rotations in the 
planes $x_1x_2$, $x_3x_4$ ,...,$x_{2p-1}x_{2p}$ 
which form the above-discussed Abelian group  $U(1)^p$.
They may be presented as explicit functions of the non-propagating 
brane coordinates $q_{a}(\sigma^r)$ and their derivatives 
 \begin{equation}\label{angmom} 
j_b=
T\omega_{b}q_{b}^2 \sqrt{\frac{|g|}{1-\sum_{a=1}^{p}\omega_{a}^2q_{a}^{2}}}\ .
\end{equation}

  We conclude that the choice of the initial value constraints in the 
form of harmonicity conditions (\ref{statn}) selects the regular or singular 
 $g^{rs}$ given by the solutions of Eqs.(\ref{xeqmp})
describing infinite or compact folded spinning $p$-branes.

\section{Solvable $p$-brane motions with $\Delta^{(p)}\vec{x}=-\Lambda\vec{x}$}

In the previous section we have  found that the harmonicity 
equations $\Delta^{(p)}\vec{x}=0$ treated as the 
 initial value constraints provide the exact 
solutions \cite{IJGM} of brane equations.
One can conjecture that specially constructed deformations 
of the harmonicity conditions may reveal other exact solutions. 
This proposal is compatible with  the specific form 
of brane Eqs.(\ref{xeqmp}), where the shift of the factor $G_{tt}$ 
to their l.h.s. leaves only $\Delta^{(p)}\vec{x}$
 in the r.h.s.. 
Therefore, the time derivatives of $\vec{x}$ are concentrated in the 
l.h.s. of (\ref{xeqmp}). Using various initial value constraints, 
including $\Delta^{(p)}\vec{x}$ in combination with $\vec{x}$ 
and $\dot{\vec{x}}$, one can generate various evolution equations. 
It may occur that some of these evolution equations are exactly 
solvable like in the case $\Delta^{(p)}\vec{x}=0$.
The constraint deformations are under control of the Noether identities 
demanding $\Delta^{(p)}\vec{x}$ 
to be an eigenvector of the projection operator $\Pi_{ik}$, 
as it follows from (\ref{lappro}). 
Variation of the constraints will result in deformations 
of the brane shape selfconsistent with the evolution equations.  
 
As an example realizing this proposal and generalizing the 
solutions \cite{Znpb} we consider the following  
 invariant deformation of the harmonicity conditions
\begin{equation}\label{genhrm}
\Delta^{(p)}\vec{x}=-\Lambda \vec{x},
\end{equation}
where $\Lambda$ is an arbitrary function invariant under 
diffeomorphisms of the hypersurface $\Sigma_{p}$. 
The substitution of (\ref{genhrm})   
into (\ref{xeqmp}) yields the evolution equation
\begin{eqnarray}\label{ekvgenhrm4} 
\ddot{\vec{x}} - (\ddot{\vec{x}}\vec{x}^{,s})\vec{x}_{,s}
=-\Lambda (1 - \dot{\vec{x}}^2)\vec{x},   \\ 
\label{ekvgenhrm1} 
(\partial_sln\sqrt{|g|})\vec{x}^{,s} + \partial_s\vec{x}^{,s}=-\Lambda\vec{x}
 \end{eqnarray}
accompanied with the constraints (\ref{ekvgenhrm1}) for the initial value for 
this evolution equation.                  
 Due to the Noether identities we obtain that the projections of 
 Eqs.(\ref{ekvgenhrm4}-\ref{ekvgenhrm1}) on $\vec{x}_{,r}$ result 
in the following rotationally invariant constraint 
\begin{equation}\label{sfera}
 \vec{x}^{2}(t,\sigma^r)=\vec{R}^2(t)
\end{equation}
which shows that the $p$-brane hypersurface $\Sigma_{p}$ resides on 
 the $(D-2)$-dimensi-onal  sphere of 
the radius $R=\sqrt{\vec{R}^2(t)}$ embedded into the $(D-1)$-dimensional
 Euclidean space. The projections of (\ref{ekvgenhrm4}-\ref{ekvgenhrm1}) 
on $\dot{\vec{x}}$  yield the equations  
\begin{eqnarray}\label{uskor}
\frac{1}{2}\frac{d^2\vec{x}^{2}}{dt^2}= 1-(p+1)(1 - \dot{\vec{x}}^2), \\
\label{otvla}
\Lambda(t,\sigma^r)=\frac{p}{\vec{R}^2(t)} \ \ \ \ \ \ \ \ \ \ \ \  
\end{eqnarray}
fixing the unknown function $\Lambda$.
The projections of (\ref{ekvgenhrm4}-\ref{ekvgenhrm1}) on $\vec{x}$ give 
the relation
\begin{equation}\label{elipt}
1 - \dot{\vec{x}}^2=(\frac{\vec{R}^2(t)}{l^2})^p,
\end{equation}
where $l$ is the integration constant with the dimension $[l]=L$. 
The latter relation in combination with (\ref{uskor}) yields 
 the closed equation for $ \xi:=\vec{R}^2(t)$
\begin{equation}\label{elipt1}
\frac{1}{2}\ddot\xi=1- (p+1)\left(\frac{\xi}{l^2}\right)^p.
\end{equation}
The first integral of Eq. (\ref{elipt1}) is given by the relation
$$
l^2\dot\zeta^{2}= (1- \zeta^{p})(1 + \zeta^{p}) \ \ \ \ \ \  
$$
expressed in terms of the new dimensionless
 variable $\zeta(t):=\frac{\sqrt{\xi}}{l}\equiv\frac{\sqrt{\vec{R}^2}}{l}$ 
 substituted instead of $\vec{R}^2(t)$.
Then the first integral is presented as
\begin{equation}\label{1stint}
(\frac{d\zeta}{d\eta})^2= \frac{1}{2}(1-\zeta^p)(1+\zeta^p)
\end{equation}
after the transition to the new rescaled time variable $\eta:=\frac{\sqrt{2}\, t}{l}$. 
For the case $p=2$ corresponding to membrane  Eq. (\ref{1stint}) 
is the defining equation for the Jacobi 
elliptic cosine $cn(\eta,k)$ with the elliptic modulus $k=\frac{1}{\sqrt{2}}$. 
For $p>2$ the exact solution of (\ref{1stint}) is given by 
the hyperelliptic integral 
\begin{equation}\label{hprelp}
\eta=\pm\sqrt{2}\int\frac{d\zeta}{\sqrt{1-\zeta^{2p}}} + const
\end{equation}
 generalizing the elliptic membrane solution to $p$-branes 
with arbitrary  $p$.
Thus, we obtain exact solution for the length $\sqrt{\vec{R}^2(t)}$ 
of $\vec{x}$ without any gauge fixing for the symmetry (\ref{diff}) 
and the restiction $D=2p+1$ \cite{Znpb}.

Then the generalized harmonicity conditions (\ref{genhrm})
take the form
\begin{equation}\label{genhrm'}
\Delta^{(p)}\vec{x}=-\frac{p}{{\vec{R}^2}(t)}\vec{x}
\end{equation} 
 with the known function $\vec{R}^2(t)$ depending only on time.
The $\sigma$-independence of $\vec{x}^2=\vec{R}^2(t)$ results in 
 the $\sigma^r$-independence of $\dot{\vec{x}}^2$, as it follows 
from (\ref{elipt}) and the fact that the second term 
in the l.h.s.  of (\ref{ekvgenhrm4}) vanishes. 
As a result, Eqs. (\ref{ekvgenhrm4}) and (\ref{ekvgenhrm1}) are reduced 
to two connected subsystems
\begin{eqnarray}\label{ekvgenhrm4*} 
\ddot{\vec{x}} + \frac{p}{l^2}(\frac{\vec{R}^2}{l^2})^{p-1}\vec{x}=0, 
\label{ekvgenhrm4'} 
\\
\Delta^{(p)}\vec{x} + \frac{p}{{\vec{R}^2}}\vec{x}=0   \label{ekvgenhrm1'} 
\end{eqnarray}
 with the evolution equations describing $2p$-dim. oscillator with 
time-dependent frequency given by the (hyper)elliptic function of time.
 To find all the components of the vector $\vec{x}$  we must 
 solve Eqs.(\ref{ekvgenhrm4'}) and (\ref{ekvgenhrm1'}).
 Since the length of $\vec{x}$ is $\sigma$-independent,  
 this dependence concentrates in the direction cosines of $\vec{x}$. 
 This suggests representation of $\vec{x}$ in the form
$
x_{i}(t, \sigma^r)=\mathcal{O}_{ik}(t, \sigma^r)R_{k}(t),
$
where $\mathcal{O}_{ik} \in SO(D-1)$ group of 
rotations of $(D-1)$-dimensional subspace of the Minkowski space.
 In view of the time independence of $|\dot{\vec{x}}|$,
 the time derivative of this representation for $\vec{x}$
  shows that the matrix $\mathcal{O}$ 
 is also time-independent. 
 This observation results in the separation of variables 
\begin{equation}\label{vrawen}
x_{i}(t, \sigma^r)=\mathcal{O}_{ik}(\sigma^r)R_{k}(t), \ \ \ 
\mathcal{O}_{ik}\mathcal{O}_{jk}=\delta_{ij}.
\end{equation} 
Similarly to the spinning brane case we restrict ourselves  by
 $(2p+1)$-dim. Minkowski space
 and choose the matrix $\mathcal{O}_{ik}$ from
 the Abelian subgroup $O(2)^p$ of the group $SO(2p)$. 
 Then $\vec{x}$ takes the form of the anzats \cite{Znpb}
 \begin{eqnarray} 
\vec{x}^T
=(q_1\cos\theta_1,q_1\sin\theta_1,q_2\cos\theta_2,q_2\sin\theta_2,
\ldots,q_p\cos\theta_p,q_p\sin\theta_p), \label{contranzats}
 \\
q_a=q_a(t),\,\,\, 
 \theta_a=\theta_{a}(\sigma^{r}). 
\, \, \, \, \,  \, \, \,\, \, \, \, \, \, \, \,\, \, \, \, \, \, \, \, \, \, \, \, 
\, \, \, \, \, \, \, \, \, \, \, \, \, \, \, \, \,  \, \, \,   \nonumber
\end{eqnarray}
Contrary to the spinning anzats (\ref{spinanzats}), 
considering its polar angles
to be propagating DOF, here we have the radial 
coordinates  $\mathbf{q}(t)=(q_1,..,q_p)$ as the propagating DOF.
Anzats (\ref{contranzats}) yields the following expressions 
for the lengths of $\vec{x}$ and $\dot{\vec{x}}$
\begin{equation}\label{norma}
\vec{x}^{2}(t,\sigma^r)=\mathbf{q}^2(t)\equiv\sum_{a=1}^{p}q^{2}_{a}(t), \ \ \ \
\dot{\vec{x}}^2(t)=\dot{\mathbf{q}}^2(t)
\end{equation}
and for the worldvolume metric $G_{\alpha \beta}$ on $\Sigma_{p+1}$, respectively
\begin{eqnarray}\label{metr} 
 G_{tt}=1- \dot{\bf{q}}^2, \  \  \  {\bf q}:=(q_1,..,q_p), \  \ \
g_{rs}= \sum_{a=1}^{p}q_{a}^{2}\theta_{a,r}\theta_{a,s},
\end{eqnarray}  
where $\theta_{a,r}\equiv \partial_{r}\theta_{a}$. 
The corresponding squared interval $ds^2_{p+1}$ is given by
\begin{equation}\label{intr*}   
ds^2_{p+1}=(1-\dot{\bf{q}}^2)dt^{2} 
- \sum_{a=1}^{p}q_{a}^{2}(t)d\theta_{a}d\theta_{a}.
\end{equation}
 
 Representation (\ref{intr*}) shows that in the new 
 coordinates $\theta_{a}(\sigma^{r})$, used instead of $\sigma^{r}$, the metric 
 on $\Sigma_{p}$ 
 becomes independent of  $\sigma^{r}$ with $p$ Killing vector fields 
 represented by the derivatives  $\frac{\partial}{\partial \theta_a}$.
Thus, anzats (\ref{contranzats}) describes p-dimensional 
torus $S^1\times S^1\times \ldots \times S^1$ with zero curvature and 
the time-dependent radii $q_{a}$.
This anzats reduces the number of degrees of freedom to $p$ carried by  
the radial coordinates $q_a$ which obey reduced Eqs.(\ref{ekvgenhrm4*})  
\begin{eqnarray}\label{ekvgenhrm4con} 
\ddot{\mathbf{q}} =-\frac{p}{l^2}(\frac{\mathbf{q}^2}{l^2})^{p-1}\mathbf{q}
\end{eqnarray}
 with their first integral equal 
to $1 - \dot{\mathbf{q}}^2=(\frac{\mathbf{q}^2(t)}{l^2})^p.$
 The substitution of expressions (\ref{norma}) in Eqs.(\ref{uskor}) 
regenerate Eq. (\ref{1stint}) and its hyperelliptic solution (\ref{hprelp})
with  $\mathbf{q}^2$ substituted for $\vec{x}^2$, 
e.d. $\eta(t)=\frac{\sqrt{\mathbf{q}^2}}{l}$.
 
The substitution of anzats (\ref{contranzats}) into Eqs. (\ref{ekvgenhrm1'}) 
 transforms them to homogeneouos equations for the components $\theta_a$  
 which are equivalent to 
 \begin{eqnarray}
 g^{rs}\theta_{a,r}\theta_{a,s}=\frac{p}{\mathbf{q}^2} \label{ekvgenhrm1a} 
\ \ \ \ (a=1,2,...,p),
 \\
 \frac{1}{\sqrt{|g|}}\partial_{r}\left( \sqrt{|g|}g^{rs}\theta_{a,s} \right)
 +  g^{rs} \theta_{a,rs}=0 \label{ekvgenhrm1b} 
 \end{eqnarray}
 for each $a$. 
 The equations are easily solved in the 
gauge $\theta_{a}=\delta_{ar}\sigma^{r}$ \cite{Znpb}
 \begin{equation}\label{qgauge}
\theta_{1}(\sigma^{r})=\sigma^{1}, \ \ \ \theta_{2}(\sigma^{r})=\sigma^{2}, \ 
 \ldots \  , \  \theta_{p}(\sigma^{r})=\sigma^{p},
\end{equation}
where $\sigma^{r}$-independent metric $g_{rs}(t)$ takes the following diagonal form 
\begin{eqnarray}\label{diago}
g_{rs}(t)= q_{r}^{2}(t)\delta_{rs}, \ \ \
 g=(q_{1} q_{2}...q_{p})^{2} \label{gmtr}  
\end{eqnarray}
and transforms Eqs. (\ref{ekvgenhrm1b}) to identities. 
Eqs. (\ref{ekvgenhrm1a}) reduce to the conditions 
\begin{eqnarray}\label{solut}
\mathbf{q}^2\equiv\sum_{a=1}^{p}q_{a}^{2}=pq_{1}^{2}=pq_{2}^{2}=...=pq_{p}^{2}
\end{eqnarray}
which mean coincidence of all $q_{a}$-functions:  $q_{a}(t)\equiv q(t)$.

From the geometrical point of view the coincidence condition picks up 
the case of degenerate $p$-torus with equal radii \cite{Znpb}.
In view of the above constraints, 
 the system of $p$ tangled equations (\ref{ekvgenhrm4con})
\begin{eqnarray}\label{ekvgenhrm4con*} 
\ddot{q}_{a} =-\frac{p}{l^2}(\frac{q_{1}^2 + q_{2}^2 + ...+ q_{p}^2}{l^2})^{p-1}q_{a} \ \ \ \
(a=1,2,...,p)
\end{eqnarray}
shrinks to the single exactly solvable nonlinear equation 
\begin{eqnarray}\label{ekvgenhrm4con''} 
\ddot{q} + \frac{p}{l^2}(\frac{pq^2}{l^2})^{p-1}q =0
\end{eqnarray}
with the above-studied first integral given by 
\begin{equation}\label{qelipt'}
1 - p\dot{q}^2=(\frac{pq^{2}}{l^2})^p.
\end{equation}
The change of variables $\tilde\zeta
=\frac{\sqrt{p}q}{l}, \, \eta=\frac{\sqrt{2}\, t}{l}$ 
transforms Eq. (\ref{qelipt'}) into  Eq. (\ref{1stint}) 
\begin{equation}\label{1stint'}
(\frac{d\tilde\zeta}{d\eta})^2= \frac{1}{2}(1-\tilde\zeta^p)(1+\tilde\zeta^p)
\end{equation}
and its solution is given by the considered 
 hyperelliptic integral $(\ref{hprelp})$
\begin{equation}\label{hprelp'}
\eta=\pm\sqrt{2}\int\frac{d\tilde\zeta}{\sqrt{1-\tilde\zeta^{2p}}} + const.
\end{equation} 

Thus, we proved that the deformation (\ref{genhrm}) of the harmonicity 
conditions selects 
the exact solution which describes collapsing $p$-brane with the shape 
of the degenerate $p$-torus \cite{Znpb}.

\section{Summary}

A new  approach to the problem of exact solvability of nonlinear
$p$-brane equations and constraints in $D$-dimensional Minkowski 
space was considered. The approach is based on the connection between 
the initial value problem for the brane equations and their exact solutions. 
  
The $p$-brane equations, initially written in the form o
f $(p+1)$-dimensional worldvolume wave equations, were reduced in 
the orthogonal gauge to $p$-dimensional equations 
with their r.h.s. presented  by $\Delta^{(p)}\vec{x}$ and l.h.s. 
equal to the brane 
acceleration projection on the directions orthogonal to its 
hypersurface $\Sigma_p$. 
The Noether identities associated with the diffeomorphisms of 
the brane worldvolume $\Sigma_{p+1}$ were derived and used for 
the choice of the admissible constraints for the initial data.   
Two types of such constraints were studied and the corresponding
 exact solutions were obtained.
  The first of them  considers the harmonicity constraints
$\Delta^{(p)}\vec{x}=0$ which select spinning $p$-branes.  
 In the case $D=2p+1$ the harmonicity constraints are exactly solved 
by the anzatz previously considered in \cite{IJGM}. 
These solutions include either regular solutions for $g^{rs}$ 
describing infinite $p$-branes with the shape of $p$-dimensional hyperplanes 
or nonregular $g^{rs}$ associated with folded compact $p$-branes.
The case of the infinite branes includes static $p$-branes with the 
constant density of energy treated as p-dimensional domain walls. 
 The second set is picked up by the deformed harmonicity conditions 
$\Delta^{(p)}=-\Lambda\vec{x}$ and describes
closed  $p$-brane lying on a collapsing sphere $S^{D-2}$
embedded into $(D-1)$-dimensional Euclidean subspace of $D$-dimensional
 Minkowski space with arbitrary $D>4$. The time-dependent radius 
of the sphere is presented by hyperelliptic functions.
 In the particular case of odd $D=2p+1$ the $p$-brane 
 hypersurface $\Sigma_{p}$ turns out to be isometric to flat collapsing 
$p$-dimensional torus which coincides with the exact solution \cite{Znpb}. 
 The described spinning or collapsing 5-branes ($p=5$) give
 exact solutions of $D=11$ M/string theory and it is interesting to
 understand the physics associated with them. 

Extension of the proposed approach to the case of 
opened $p$-branes with various boundary conditions
as well as its generalization to the case of
 known cosmological backgrounds seems to be interesting.

\noindent{\bf Acknowledgments}

The author is grateful to Physics Department 
of Stockholm University and 
 Nordic Institute for Theoretical  NORDITA 
for kind hospitality and financial support.
The results  were presented at the Conference STDE-2012 
in honor of Vladimir Aleksandrovich Marchenko's 90th birthday \cite{VAM}.

\end{document}